\documentclass[nofootinbib,prd,showpacs]{revtex4}
\usepackage{amsmath}
\usepackage{amsfonts}
\usepackage{amssymb}
\usepackage{graphicx}

\begin{document}

\title{Exact solutions of Brans-Dicke wormholes in the presence
of matter}

\author{Nadiezhda Montelongo Garcia}
\email{nadiezhda@cosmo.fis.fc.ul.pt}\affiliation{Centro de Astronomia
e Astrof\'{\i}sica da Universidade de Lisboa, Campo Grande, Edif\'{i}cio C8
1749-016 Lisboa, Portugal}

\author{Francisco S.~N.~Lobo}
\email{flobo@cii.fc.ul.pt}\affiliation{Centro de Astronomia
e Astrof\'{\i}sica da Universidade de Lisboa, Campo Grande, Edif\'{i}cio C8
1749-016 Lisboa, Portugal}

\date{\today}

\begin{abstract}

A fundamental ingredient in wormhole physics is the presence of exotic matter, which involves the
violation of the null energy condition. Although a plethora of wormhole solutions have been explored
in the literature, it is useful to find geometries that minimize the usage of exotic matter. In this
work, we find exact wormhole solutions in Brans-Dicke theory where the normal matter threading the
wormhole satisfies the null energy condition throughout the geometry. Thus, the latter implies that
it is the effective stress-energy tensor containing the scalar field, that plays the role of exotic 
matter, that is responsible for sustaining the wormhole geometry. More specifically, we consider a 
zero redshift function and a particular choice for the scalar field and determine the remaining 
quantities, namely, the stress-energy tensor components and the shape function. The solution found 
is not asymptotically flat, so that this interior wormhole spacetime needs to be matched to an 
exterior vacuum solution.

\end{abstract}


\maketitle

\section{Introduction}

Wormholes are hypothetical short-cuts in spacetime and, in classical general relativity, are
supported by exotic matter \cite{Morris}. The latter involves a stress-energy tensor $T_{\mu\nu}$
that violates the null energy condition (NEC), i.e., $T_{\mu\nu}k^{\mu}k^{\nu}<0$, where $k^{\mu}$
is a null vector. In fact, wormholes violate all the pointwise energy conditions and the averaged
energy conditions. The latter averaged energy conditions are somewhat weaker than the pointwise
energy conditions, as they permit localized violations of the energy conditions, as long on average
the energy conditions hold when integrated along timelike or null geodesics \cite{Tipler}. In this
context, although a plethora of wormhole solutions have been explored in the literature
\cite{whsolutions}, it is useful to find geometries that minimize the usage of exotic matter and
this has been obtained in several manners. In \cite{Visser:2003yf} a suitable measure for
quantifying this notion was developed, and it was demonstrated that spacetime geometries containing
traversable wormholes supported by arbitrarily small quantities of ``exotic matter'' can indeed be
constructed. Another way to minimize the usage of exotic matter is to construct thin-shell wormholes
using the cut-and-paste procedure, where the exotic matter is concentrated at the throat
\cite{thinshell}. In the context of modified theories of gravity, it has also been shown that the
stress-energy tensor can be imposed to satisfy the NEC, and it is the higher order curvature terms,
interpreted as a gravitational fluid, that sustain these non-standard wormhole geometries,
fundamentally different from their counterparts in general relativity \cite{modgrav}.

In this context, we find exact wormhole solutions in Brans-Dicke theory where the normal matter
threading the wormhole satisfies the NEC, and it is the effective stress-energy tensor containing
the scalar field that is responsible for the null energy condition violation. The analysis in this
paper builds on previous work \cite{Anchordoqui:1996jh}. In the latter work, the authors consider an
equation of state give by $p_r+p_t=\epsilon \rho$, where $p_r$, $p_t$ and $\rho$ are the radial
pressure, tangential pressure and the energy density, respectively. The authors then deduce the form
of the scalar field and outline rather general conditions of the parameter space in which the Brans-
Dicke field may play the role of exotic matter, implying that it might be possible to build a
wormholelike spacetime with the presence of ordinary matter at the throat. We generalize the latter
work, showing that it is possible to construct wormhole geometries with normal matter satisfying the
NEC throughout the spacetime. These works, in turn, generalize the vacuum Brans-Dicke wormholes that
have been analysed in the literature
\cite{Agnese:1995kd,Anchordoqui:1996jh,Lobo:2010sb,Nandi:1997en,Nandietal,Nandietal2}. In
particular, specific solutions were constructed showing that static wormhole solutions in vacuum
Brans-Dicke theory only exist in a narrow interval of the coupling parameter \cite{Nandi:1997en},
namely, $-3/2<w<-4/3$. It should be emphasized that this range is obtained for vacuum solutions and
for a specific choice of an integration constant of the field equations given by $C(w)=-1/(w+2)$.
The latter relationship was derived on the basis of a post-Newtonian weak field approximation, and
there is no reason for it to hold in the presence of compact objects with strong gravitational
fields. In fact, the choice given by the above form of $C(w)$ is a tentative example and reflects
how crucially the wormhole range for $w$ depends on the form of $C(w)$. Indeed, specific examples
for $C(w)$ were given in \cite{Lobo:2010sb} that lie outside the range $-3/2<w<-4/3$.

This paper is organised in the following manner: In Section \ref{Sec:II}, we write the field
equations and the specific constraints imposed by the violations of the null energy condition. In
Section \ref{Sec:III}, we consider a specific solution by considering a zero redshift function and a
particular choice for the scalar field. Finally, in Section \ref{Sec:Conclusion}, we conclude.

\section{Field equations}\label{Sec:II}

The field equations of Brans-Dicke theory are given by \cite{bruck}
\begin{eqnarray}
R_{\mu\nu} & = & \frac{8\pi}{\Phi}\left( T_{\mu\nu}- \frac{\omega
+1}{2\omega +3}\;T\: g_{\mu\nu}\right) + \omega\:
\frac{\Phi_{;\mu}\Phi_{;\nu}}{\Phi^2}
+ \frac{\Phi_{;\mu;\nu}}{\Phi}\,, \\
\Phi^{;\mu}_{\; ;\mu} & = & \frac{8\pi}{ 2\omega +3}\; T \,,
\end{eqnarray}
where $T$ is the trace of the stress-energy tensor, $\Phi$ is the scalar field, $\omega$ the Brans-
Dicke parameter, $R_{\mu\nu}$ the Ricci tensor and $g_{\mu\nu}$ the metric tensor.

Using the the line element written in Schwarzschild coordinates
\begin{equation}
ds^2 = -e^{2\psi}dt^2 + e^{2\lambda} dr^2 + r^2 ( d\theta^2 +
\sin^2\theta d\phi ^2) \label{met} \,,
\end{equation}
the field equations take the following form
\begin{eqnarray}
-\psi''- (\psi')^2 + \lambda'\psi'+ 2 \frac{\lambda'}{r} &=&
\frac{8\pi}{\Phi}\left[p_r -\frac{\omega +1}{2\omega +3}\;T\right]
e^{2\lambda}+(\omega +1) (\ln\Phi)'^2 + (\ln\Phi)''-\lambda'
(\ln\Phi)' \label{fieq1}\,,\\
 1 - r e^{-2\lambda}\left[\psi '
- \lambda ' + \frac{1}{r}\right]  &=& \frac{8\pi}{\Phi}\left[ p_{t} -
\frac{\omega +1}{2\omega +3}\;T\right]r^2 + r e^{-2\lambda}(\ln
\Phi)' \label{fieq2}\,,\\
e^{2(\psi -\lambda)}\left[\psi '' + (\psi ')^2 - \lambda ' \psi '
+ 2 \frac{\psi '}{r}\right]  &=& \frac{8\pi}{\Phi}\left[\rho +
\frac{\omega +1}{2\omega +3}\;T\right] e^{2\psi} - \psi '
e^{2(\psi - \lambda)} (\ln \Phi)' \label{fieq3} \,, \\
\Phi '' - \Phi ' \left( \lambda ' - \psi ' - \frac{2}{r}\right)
&=& \frac{8\pi}{2\omega + 3}\; T \; e^{2\lambda} \label{fieq4} \,,
\end{eqnarray}
respectively, where the prime denotes a derivative with respect to the radial
coordinate. $\rho(r)$ is the energy density, $p_r(r)$ is the
radial pressure, and $p_t(r)$ is the tangential pressure.
The Bianchi identity, $G^{\mu\nu}{}_{;\nu}=0$, implies that for a static spherically
symmetric anisotropic fluid, we have
\begin{eqnarray}
p_{r}'=-(\rho+p_{r})\psi'+\frac{2(-p_{r}+p_{t})}{r}\label{eq3} \,.
\end{eqnarray}

One now has at hand four equations, namely, the field Eqs. (\ref{fieq1})-(\ref{fieq4}), with six
unknown functions of $r$, i.e., $\rho(r)$, $p_r(r)$, $p_t(r)$, $\psi(r)$, $\lambda(r)$ and
$\Phi(r)$. To construct specific solutions, one may adopt several approaches, and in this work we
mainly use the strategy of considering restricted choices for the redshift function and for
the scalar field $\Phi(r)$, in order to obtain solutions with the properties and characteristics of
wormholes.

A fundamental property of wormhole physics is the violation of the null energy condition at the
throat \cite{Morris}. In this context, it is useful to write the field equation in the form
$G_{\mu\nu}=8\pi T_{\mu\nu}^{\textrm{Eff}}$ where $T_{\mu\nu}^{\textrm{Eff}}$ is given by
\begin{eqnarray}
T_{\mu\nu}^{Eff}=
\frac{T_{\mu\nu}}{\Phi}+\frac{1}{8\pi}\left[  \frac{\omega}{\Phi^{2}}
(\Phi_{;\mu}\Phi_{;\nu}-\frac{1}{2}g_{\mu\nu}\Phi ^{;\alpha}\Phi _{;\alpha})
+\frac{1}{\Phi}(\Phi_{;\mu\nu}-g_{\mu\nu}
\Phi ^{;\alpha}_{;\alpha}) \right]\,.
\end{eqnarray}
The violation of the NEC states that $T^{\textrm{Eff}}_{\mu\nu}k^{\mu}k^{\nu}<0$, where $k$ is {\it
any} null vector, and imposes the following constraint
\begin{eqnarray}
\left\{[8\pi(\rho +p_{r})+
e^{-2\lambda}\left[\Phi''-\Phi'(\lambda'+\Psi')
+\frac{\omega}{\Phi}(\Phi')^{2}\right]\right\}\Big |_{r_{0}}<0\,.\label{C1}
\end{eqnarray}

In this work, we also impose that the normal matter threading the wormhole satisfies the NEC,
$T_{\mu\nu}k^{\mu}k^{\nu}>0$. At the throat this states that $(\rho +p_{r})|_{r_{0}}>0$.
Thus, it is useful to rewrite the components $\rho(r)$, $p_r(r)$ and $p_t(r)$, as a function
of the metric functions $\psi(r)$ and $ \lambda(r)$ and the scalar field $\Phi(r)$. Note that one
may simply substitute the factor $T/(2\omega +3)$ of Eq. (\ref{fieq4}) into Eqs.
(\ref{fieq1})-(\ref{fieq3}), which yield the following relationships
\begin{eqnarray}
\rho&=&\frac{e^{-2\lambda}\Phi}{8\pi}\left\{\psi''+\psi'^{2}+\psi'\left[-\lambda'
-\omega(\ln\Phi)'+\frac{2}{r}\right]-(\omega+1)\left[(\ln\Phi)''
+(\ln\Phi)'-(\ln\Phi)'\left(\lambda'-\frac{2}{r}\right)\right]\right\},\\
p_{r}&=&\frac{e^{-2\lambda}\Phi}{8\pi}\left\{-\psi''-\psi'^{2}+\psi'[\lambda'
+(\omega+1)(\ln\Phi)']+\omega(\ln\Phi)''
+(\ln\Phi)'(\omega\lambda'+\frac{2}{r}(\omega+1))\right\},\\
p_{t}&=&\frac{\Phi}{8\pi r^{2}}\left\{1-re^{-2\lambda}(\psi'-\lambda'+(\ln\Phi)'+\frac{1}{r})
+(\omega+1)r^{2}
\left[(\ln\Phi)''+(\ln\Phi)'2-(\ln\Phi)'(\lambda'-\psi'-\frac{2}{r})\right]\right\}\,,
\end{eqnarray}
respectively. These equations will be explored further below.

\section{Exact solution: Zero redshift function}\label{Sec:III}

In this section, we consider the wormhole metric given in the more familiar spherically symmetric
and static form \cite{Morris}
\begin{equation}
ds^2=-e ^{2\psi(r)}\,dt^2+\frac{dr^2}{1- b(r)/r}+r^2 \,(d\theta
^2+\sin ^2{\theta} \, d\phi ^2) \label{metricwormhole}\,.
\end{equation}
Comparing with the metric (\ref{met}), this corresponds to the following identification
\begin{equation}
\lambda(r)=-\frac{1}{2}\ln \left(1- \frac{b(r)}{r}\right)\,,
\end{equation}
where $\psi(r)$ and $b(r)$ are arbitrary functions of the radial coordinate, $r$. Following the
conventions used in \cite{Morris}, $\psi(r)$ is denoted as the redshift function, for it is related
to the gravitational redshift; $b(r)$ is called the shape function, because as can be shown by
embedding diagrams, it determines the shape of the wormhole \cite{Morris}. A fundamental property of
a wormhole is that a flaring out condition of the throat, given by $(b-b'r)/b^2 > 0$, is imposed
\cite{Morris}, and at the throat $b(r_0) = r = r_0$, the condition $b'(r_0) < 1$ is imposed to have
wormhole solutions. This latter condition will be explored below. Note that the condition $1-b/r >
0$ is also imposed. It is possible to construct asymptotically flat spacetimes, in which $b(r)/r
\rightarrow 0$ and $\psi \rightarrow 0$ as $r \rightarrow \infty$. However, one may also construct
solutions by matching the interior solution to an exterior vacuum spacetime, at a junction
interface, much in the spirit of \cite{matching}. The latter case is applied in this work, as will
be shown below.

For simplicity, we consider the specific case of a zero redshift function, $\psi=0$, so that Eq.
(\ref{fieq3}) provides the following relationship
\begin{eqnarray}
p_{t}=-\frac{1}{2}\left[p_{r}+\left(\frac{\omega+2}{\omega+1}\right)\rho\right],
\end{eqnarray}
and (\ref{eq3}) yields
\begin{eqnarray}
p_{r}'=-\frac{1}{r}\left[3p_{r}
+\left(\frac{\omega+2}{\omega+1}\right)\rho\right].
\end{eqnarray}

In addition to the zero redshift function, we consider the ansatz $\Phi=\Phi_{0}\left(\frac{r_{0}}
{r}\right)^{\alpha},$ with
$\Phi_{0}>0$ and $\alpha>0$, so that the field equations can be written in the following manner
\begin{eqnarray}
\left(1-\frac{\alpha}{2}\right)b'
+\left[(\omega+1)\alpha^{2}+\frac{3\alpha}{2}-1\right]\frac{b}{r}
-(\omega+1)\alpha^{2}-\alpha
&=&\frac{8\pi r^{2}}{\Phi_{0}}(\rho+p_{r})\left(\frac{r_{0}}{r}\right)^{-\alpha},\\
b'+\frac{b}{r}(1-2\alpha)+2\alpha &=&\frac{8\pi r^{2}}{\Phi_{0}}\left[\left(\frac{\omega}
{\omega+1}\right)\rho-p_{r}\right]
\left(\frac{r_{0}}{r}\right)^{-\alpha},\\
b'+(1-2\alpha)\frac{b}{r}+2(\alpha-1)
&=&-\frac{16\pi\rho r^{2}}{(\omega+1)\alpha\Phi_{0}}
\left(\frac{r_{0}}{r}\right)^{-\alpha}.
\end{eqnarray}

The solution of this system is given by the following stress-energy components
\begin{eqnarray}
\rho(r)&=&\frac{1}{4\pi r_0^2}\left[\frac{\alpha(\omega+1)c_{1}}{\omega\alpha+2}
\left(\frac{r_0}{r}\right)^{\frac{2(\alpha+2\omega\alpha+3)}{\omega\alpha+2}}
+\frac{(\alpha-1)\alpha\Phi_{0}(\omega+1)}{(\omega\alpha^{2}-2\omega\alpha-2)}\left(\frac{r_{0}}
{r}\right)^{\alpha+2}\right],\label{rho} \\
p_{r}(r)&=&\frac{1}{4\pi r_0^2}\left[c_1\left(\frac{r_0}{r}\right)^{\frac{2(\alpha+2\omega\alpha+3)}
{\omega\alpha+2}}
+\frac{(\omega+2)\alpha\Phi_{0}}{(\omega\alpha^{2}-2\omega\alpha-2)}
\left(\frac{r_{0}}{r}\right)^{\alpha+2}\right],\label{pr}  \\
p_{t}(r)&=&-\frac{1}{4\pi r_0^2}\left[\frac{c_{1}(\omega\alpha+\alpha+1)}{\omega\alpha+2}
\left(\frac{r_0}{r}\right)^{\frac{2(\alpha+2\omega\alpha+3)}{\omega\alpha+2}}
+\frac{\alpha^{2}\Phi_{0}(\omega+2)}{2(\omega\alpha^{2}-2\omega\alpha-2)}\left(\frac{r_{0}}
{r}\right)^{\alpha+2}\right]\,,\label{pt}
\end{eqnarray}
respectively, and the shape function is provided by
\begin{eqnarray}
b(r)=r_0\left[\frac{\omega\alpha(\alpha-2)}{\omega\alpha^{2}-2\omega\alpha-2}\left(\frac{r}
{r_0}\right)+\frac{4 c_{1}}{\Phi_{0}(\omega\alpha^{2}+4\alpha-2)}
\left(\frac{r_{0}}{r}\right)^{\frac{\alpha\omega(1-\alpha)}{\omega\alpha+2}}
+c_{2}\left(\frac{r_0}{r}\right)^{-1+2\alpha}\right],\label{shape-b}
\end{eqnarray}
where $c_1$ and $c_2$ are constants of integration. $c_1$ is given by
\begin{eqnarray}
c_{1}=\frac{(\omega\alpha+2)[4\pi r_{0}^{2}(\omega\alpha^{2}-2\omega\alpha-2)(\rho_{0}+p_{r}(r_{0}))
-(\omega\alpha^{2}+\alpha^{2}+\alpha)\Phi_{0}]}{(\omega\alpha^{2}-2\omega\alpha-2)
(\alpha+2\omega\alpha+2)}. \label{const1}
\end{eqnarray}
One may determine $c_2$ through the condition $b(r_{0})=r_{0}$, which yields
\begin{eqnarray}
c_{2}=-\frac{4\pi c_{1}}{\Phi_{0}(\omega\alpha^{2}+4\alpha-2)}
-\frac{2}{(\omega\alpha^{2}-2\omega\alpha-2)}.
\end{eqnarray}

For simplicity, consider the specific case $c_{1}=0$, which from Eq. (\ref{const1}) imposes the
condition
\begin{equation}
\rho(r_0)+p_r(r_0)=\frac{\alpha\Phi_0\left[1+\alpha(1+\omega)\right]}{4\pi
r_{0}^2(\omega\alpha^{2}-2\omega\alpha -2)}\,,
\end{equation}
and the field equations (\ref{rho})-(\ref{pt}) reduce to the following relationships
\begin{eqnarray}
\rho(r)&=&\frac{1}{4\pi r_0^2}
\frac{\alpha\Phi_{0}(\alpha-1)(\omega+1)}{(\omega\alpha^{2}-2\omega\alpha-2)}\left(\frac{r_{0}}
{r}\right)^{\alpha+2},\label{rho0}\\
p_{r}(r)&=&\frac{1}{4\pi r_0^2}
\frac{\alpha(\omega+2)\Phi_{0}}{(\omega\alpha^{2}-2\omega\alpha-2)}\left(\frac{r_{0}}
{r}\right)^{\alpha+2},\label{pr0}\\
p_{t}(r)&=&-\frac{1}{4\pi r_0^2}
\frac{\alpha^{2}(\omega+2)\Phi_{0}}{2(\omega\alpha^{2}-2\omega\alpha-2)}\left(\frac{r_{0}}
{r}\right)^{\alpha+2}\,.\label{pt0}
\end{eqnarray}
The shape function takes the form
\begin{eqnarray}
b(r)=\frac{\omega\alpha(\alpha-2)r}{\omega\alpha^{2}-2\omega\alpha-2}
\left[1-\frac{2}{\omega\alpha(\alpha-2)}\left(\frac{r_0}{r}\right)^{2\alpha}\right]
\,. \label{shape2}
\end{eqnarray}
In the general case, and despite the fact that the stress-energy tensor profile tends to zero at
infinity for $\alpha>0$, the wormhole solutions considered in this work are not asymptotically flat,
as can be readily verified form the Eq. (\ref{shape2}). However, for these cases, one matches an
interior wormhole solution with an exterior vacuum Schwarzschild spacetime, much in the spirit of
\cite{matching}.

The flaring-out condition of the throat, which implies the fundamental condition
$b'(r_0) < 1$, is imposed to have wormhole solutions and for the present solution is given by:
\begin{eqnarray}
b'(r_{0})=\frac{\omega\alpha^{2}-2\omega\alpha-4\alpha+2}{\omega\alpha^{2}-2\omega\alpha-2}<1\label{ci},
\end{eqnarray}
Note that the restriction $\omega\alpha^{2}-2\omega\alpha-2\neq 0$ or $\omega\neq
2/[\alpha(\alpha-2)]$ is imposed. We separate the cases $0<\alpha<2$ and $2<\alpha$.

\begin{figure}[ht]
  \centering
  \includegraphics[width=3.5in]{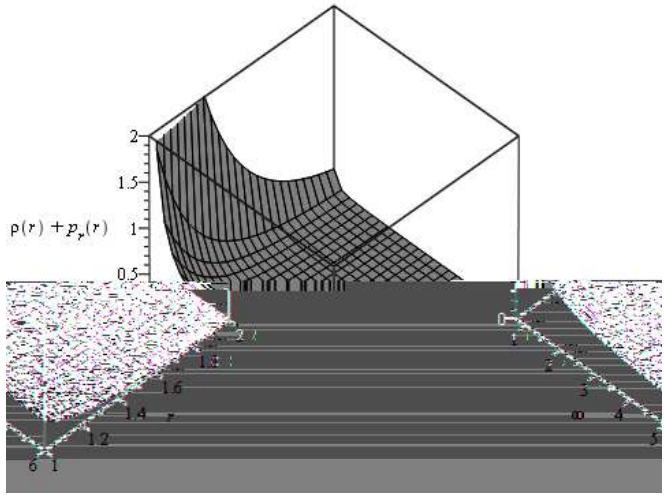}
   \caption{The plot shows that normal matter satisfies the NEC, i.e., $\rho(r)+p_r(r)>0$. The
   specific values $\Phi_{0}=1$, $r_{0}=1$ and $\alpha=3$ are considered. Note that for large values
   of $r$ and of $\omega$ the NEC tends to zero. See the text for details.}
  \label{fig1}
\end{figure}

In addition to the fundamental conditions $b'(r_0)<1$ and $r-b(r)>0$, we also impose that the normal
matter threading the wormhole obeys the NEC. It is possible to show that for the range $0<\alpha<2$,
normal matter violates the NEC, i.e., $\rho(r)+p_r(r)<0$ for all values of $r$. Thus, the only
case of interest is $\alpha>2$, for which normal matter satisfies the NEC. This is depicted in Fig.
\ref{fig1}, for the specific values of $\Phi_{0}=1$, $r_{0}=1$, $\alpha=3$. Note that for large
values of $r$ and of $\omega$ the NEC tends to zero, as can also be readily verified from Eqs.
(\ref{rho0})-(\ref{pr0}).

\begin{figure}[ht]
  \centering
  \includegraphics[width=3.5in]{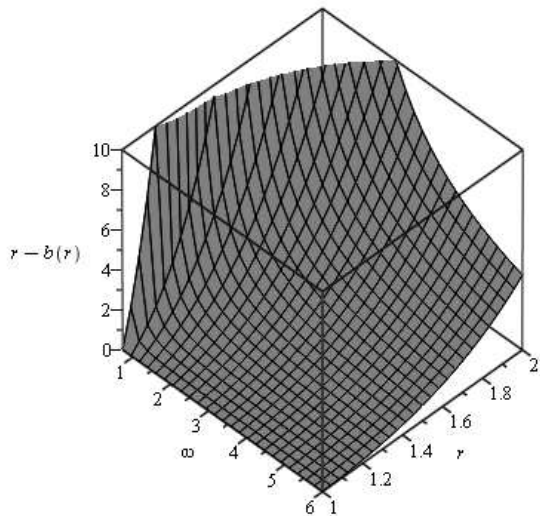}
  \caption{The plot depicts the condition $r-b>0$, with the following values: $r_{0}=1$ and $
  \alpha=3$.}
  \label{fig2}
\end{figure}

\begin{figure}[ht]
  \centering
  \includegraphics[width=3.5in]{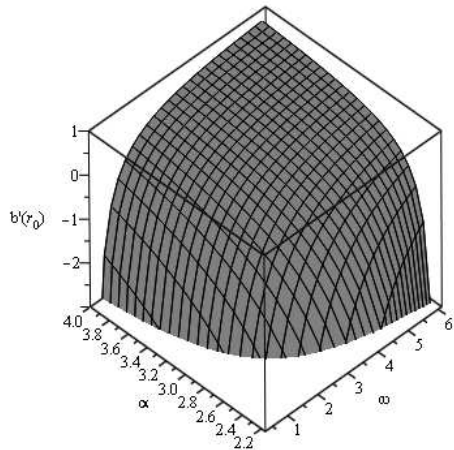}
  \caption{The plot depicts the condition for $b'<1$.  Note that for large values of $\omega$ and
  $\alpha$ we have $b'(r_0) \approx 1$, so that the throat flares out very slowly. See the text for
  details.}
  \label{fig3}
\end{figure}

The condition $r-b(r)>0$ is depicted in Fig. \ref{fig2}, also for the specific values of
$\Phi_{0}=1$, $r_{0}=1$, $\alpha=3$. The flaring-out restriction, $b'(r_0)<1$ is depicted in Fig.
\ref{fig3} for $r_0=1$. Note that for large values of $\omega$ and $\alpha$ we have $b'(r_0) \approx
1$, so that the throat flares out very slowly.

\section{Conclusion}\label{Sec:Conclusion}

A fundamental ingredient in wormhole physics is the presence of exotic matter, which involves a
stress-energy tensor that violates the null energy condition. A wide range of the solutions have
been found in the literature that minimize the usage of exotic matter such as thin shell wormholes,
where using the cut-and-paste procedure the exotic matter is minimized and constrained to the
throat. In the context of modified theories of gravity, it has also been shown that one can impose
that the normal matter satisfies the null energy condition and it is the effective stress-energy
tensor, containing higher curvature terms which can be interpreted as a gravitational fluid, that
sustain the wormhole geometry.

In this work, we have also found exact wormhole solutions in Brans-Dicke theory where the normal
matter threading the wormhole satisfies the null energy condition for a determined parameter range.
More specifically, we have imposed a zero redhsift function more computational simplicity and
imposed a specific form for the scalar field. We have found that it is the scalar field that plays
the part of exotic matter these solutions, which is in clear contrast to the classical general
relativistic static wormhole solutions. It would be interesting to analyse time-dependent solutions
and work along these lines is in progress.

\section*{Acknowledgments}
NMG acknowledges financial support from CONACYT-Mexico. FSNL acknowledges financial support of the
Funda\c{c}\~{a}o para a Ci\^{e}ncia e Tecnologia through Grants PTDC/FIS/102742/2008 and
CERN/FP/116398/2010.

\end{document}